\documentclass[twocolumn,longbibliography]{revtex4-2}
\usepackage{url}
\usepackage{color}
\usepackage{graphicx}
\usepackage{amsfonts}
\usepackage[]{natbib}
\usepackage{booktabs}
\usepackage{amsmath}
\usepackage{setspace}
\usepackage{etoolbox}
\usepackage{xcolor}
\usepackage{color,soul}
\usepackage{mathtools}
\usepackage{afterpage}

\DeclareMathOperator{\Tr}{Tr}

\DeclareMathOperator{\diag}{diag}

\usepackage{float}
\usepackage{color}
\usepackage{multirow}
\usepackage{amsmath,amssymb,amsthm}
\usepackage{graphicx}
\usepackage{setspace}
\usepackage{booktabs}
\usepackage{epstopdf}
\usepackage{footnote}
\usepackage[footnotesep=0.8\baselineskip]{geometry}
\usepackage{rotating}
\usepackage[compact]{titlesec}
\usepackage{bm}
\usepackage{lipsum} 
\usepackage{mathtools}
\usepackage{fancyhdr}
\usepackage{url}
\usepackage[hidelinks]{hyperref}
\usepackage{booktabs}
\usepackage{xcolor}
\makesavenoteenv{tabular}
\usepackage[bottom]{footmisc}
\usepackage{fmtcount} 
\usepackage{srcltx} 
\usepackage{microtype}
\usepackage{array}
\usepackage{capt-of}
\usepackage{caption}
\usepackage{subcaption}
\usepackage{tablefootnote}
\usepackage{comment}
\newcommand{\vectornorm}[1]{\left|\left|\right|\right|}
\makeatletter
\long\def\@makefigcaption#1#2{%
\vskip\abovecaptionskip
\sbox\@tempboxa{\textbf{#1 #2}}%
\global \@minipagefalse
\hb@xt@\hsize{\hfil\box\@tempboxa\hfil}%
\vskip\belowcaptionskip}
\long\def\@maketblcaption#1#2{%
\vskip\abovecaptionskip
\sbox\@tempboxa{\textbf{#1 #2}}%
\global \@minipagefalse
\hb@xt@\hsize{\hfil\box\@tempboxa\hfil}%
\vskip\belowcaptionskip}
\long\def\symbolfootnote[#1]#2{\begingroup%
\def\thefootnote{\fnsymbol{footnote}}\footnote[#1]{#2}\endgroup}

\fancypagestyle{plain}{%
\fancyhf{} 
\fancyhead[R]{\includegraphics[scale=0.2]{AuxFiles/Fulcrum_Teal-Blue.png}}

}

\definecolor{fulcrumTeal}{RGB}{122, 202, 197}
\definecolor{fulcrumBlue}{RGB}{025, 057, 071}

\titleformat{\section}{\color{fulcrumBlue!100}\bf\sffamily}{\color{fulcrumBlue!100}\thesection.}{2pt}{}

\titleformat{\subsection}{\color{fulcrumBlue!100}\large\sffamily}{\color{fulcrumBlue!100}\thesubsection.}{2pt}{}

\renewcommand{\thesection}{\arabic{section}}

\begin{document}

\title{Notes on Correlation Stress Tests }
\author{Piotr Chmielowski\\
Fulcrum Asset Management LLP\\
\today
}

\begin{abstract}
This note outlines an approach to stress testing of covariance of financial time series, in the context of financial risk management. It discusses how the geodesic distance between covariance matrices  implies a notion of plausibility of covariance stress tests. In this approach, correlation stress tests span a submanifold of constant determinant of the Fisher--Rao manifold of covariance matrices. A parsimonious geometrically invariant definition of arbitrarily large correlation stress tests is proposed, and a few examples are discussed. 
\end{abstract}

\maketitle

\newpage
\fancyhead[L]{\textsc{Notes on Correlation Stress Tests}}   

\section{Introduction} \label{sec:introduction}

Financial market news sometimes mention changes of correlations between  assets in articles with headlines such as: ``Correlations Between Credit and Equities Are Breaking Down''\cite{bloomberg_cor_breaks},``Negative equity/bond correlation is positive for 60-40 portfolio''\cite{reuters_cor_breaks}, or ``Rates correlations break down amid volatility surge''\cite{risk_magazine_stress}. Traders and asset managers also refer to correlation breaks, sometimes in a language that implies such breaks are of special interest. In this note we consider several quantitative aspects of  changes in correlations and present them in a form suitable for quantitative risk management of financial investments.

What is usually referred to as a ``correlation break" is often a  simultaneous move of the time series of several markets (in this note also referred to as market factors) that is surprising, given their past covariance.  This is likely to be an intuitive assessment of the Mahalanobis distance $d$ from zero of the vector $x$ of the recent market changes with respect to their forecasted covariance matrix $\Sigma$, namely a move which results in a large increase of 
\begin{equation} \label{lab:mahalanobis}
    d^2 = x^{T} \Sigma^{-1} x.
\end{equation}

However, there is also  another notion of  a ``correlation break", namely that of realised correlations between markets being meaningfully different from their forecasted correlations.

These two notions are related: if the former type of correlation breaks keep recurring for an extended time, they may then result in the latter. In this note, we will  consider the latter type of breaks, and will model such events as a change in the forecasted covariance matrix.

Stress tests are supposed to be unlikely, extreme, but plausible \cite{BoE_supervisory_ccp,mit_stress_test,imf_stress}. In our context, ``unlikely" probably refers to the low likelihood of a certain covariance matrix given the past data, while ``extreme" is bound to mean a large and negative impact on the entity under consideration, in our case a portfolio of financial assets or derivatives. 
The meaning of plausibility is however less clear. Various approaches to designing and selecting stress tests with reference to the empirical time series of market factors have been proposed. For example, Ref. \cite{j1996riskmetrics} describes what is perhaps the most common, albeit somewhat arbitrary, practice of stress testing as applied by investment risk management professionals, while  Ref. \cite{packham2022correlationscenarioscorrelationstress} describes stress testing of the correlation matrix of asset returns in a specific parametric form, where correlations are represented as a function of risk factors, such as country and industry factors, and where a sparse factor structure linking assets and risk factors is built using Bayesian variable selection methods. 

In the macroeconomic context,  Antolin-Diaz et al. \cite{juan_2021} developed tools for constructing econometrically meaningful scenarios with structural vector autoregressive models, and proposed a quantitative measure to assess and compare their plausibility, given the macroeconomic data. 

In this note we will consider some general aspects that abstract away from the specific interpretation of the underlying data, and concentrate instead of the geometric structure of stress tests. Thus, the methods and measures described here may be applied generically to many problems where stress testing, and particularly correlation stress testing, is of interest. While the approach is generic, we will illustrate the main conclusions with a few example applications.

The Mahalanobis distance provides a useful hint regarding the plausibility of a stress test: perhaps refers to $d$ being relatively large \cite{mit_stress_test}. For the Fisher--Rao manifolds of the multivariate normal probability  distributions, the Mahalanobis distance is a special case of the geodesic distance between the distributions: the case when they have different means, but the same covariance. The opposite case, where the multivariate distributions have  the same means, but different covariances, will be considered here. 

It is intuitively clear what a correlation stress should be: one would like to keep all the variances fixed and change the correlations -- ideally just a few or perhaps even just one correlation coefficient -- and recombine the so-stressed correlation matrix with the unchanged variances to produce the stressed covariance matrix. This however has a few drawbacks. It can trivially result in a non-positive covariance matrix. Moreover, it may fail to be consistent with no-arbitrage constraints that are present in the markets. However, by far the main drawback of such an approach is its arbitrariness. If one used a different set of factors to describe the exposure of a portfolio, an entirely different correlation stress test would need to be specified. Presumably, the specification of a stress test would be tailored to market factors specified for a portfolio, thus such a correlation stress test would depend implicitly on the portfolio being considered, and also on the somewhat arbitrary selection of the preferred market variables, or market factors, used to define it. Needless to say, it is unlikely that a universal measure of plausibility can be associated with such an arbitrary notion of stress tests. 

A change of the risk factors used to describe the portfolio can be thought of as a change of basis of the vector space of market risk factors, at least in the leading order of approximation. Requiring invariance under a change of  basis in the vector space of the portfolio risk factors may not appear important at first, but it is essential - a meaningful stress test needs to have the same result when, for example,  a fixed-income portfolio that is being stressed is represented using forward rates or alternatively represented using par rates. It is also crucial when some market variables, as a result of no-arbitrage conditions, are related through algebraic identities, as is the case for log-returns of foreign exchange rates. For example, the log-return of the USD/JPY exchange rate added to the log-return of the JPY/GBP exchange rate should be equal  to the log-return of the USD/GBP exchange rate, within the accuracy of foreign-exchange arbitrage. Projecting the stress tests on various subspaces of the space of market variables should have consistent results and a measure of the plausibility of a stress test should not depend on the choice of market variables of convenience. 

In what follows we propose an approach to defining arbitrarily large correlation stress tests. The approach uses the geometric structure of the space of correlation stress tests, which turns out to form a geodesically complete sub-manifold of the Fisher--Rao manifold of multivariate normal distributions. As a result, there is a natural measure of the size of the stress, as well as a measure of its plausibility.  The general conclusions are independent of the selection of particular coordinates on the Fisher--Rao manifold, and thus are geometrically invariant. In particular, they describe stress tests that are independent of the holdings of a portfolio or the representation of the risk of these holdings when expressed using different market factors. This is useful when one aims to apply stress tests consistently across  different portfolios. The proposed approach ends up having three useful features. First, it is exhaustive, in the sense that it allows one to explore all correlation stress tests. Second, it is universal, in the sense that the same stress tests can be applied to any covariance matrix. Third, it is quantifiable, in the sense that it provides a natural measure of plausibility of a stress test. 

This note is organised as follows: the next section introduces certain well-known features of the Fisher--Rao manifold. 
Section \ref{section:correlation_stress_tests} discusses the meaning of covariance and correlations stress tests and Section \ref{section:geodesics} discusses the interpretation of correlation stress tests as geodesics in the Fisher--Rao manifold. 
Section \ref{section:parametrisation} proposes an explicit parametrisation of the correlation stress tests which is the exponential map of the tangent space at the un-stressed covariance. Section \ref{section:plausibility} discusses the plausibility of the stress tests, while sections \ref{section:single_pair_stress} through \ref{section:all_correlations_of_single_variable} discuss the stress tests that result from specific choices of  ``directions'' in which the correlation matrix is stressed. 
In Section \ref{section:bias_of_eigenvalues} we show that a correlation stress test which affects all correlations by the same amount results in  the same lowest-order functional form as the estimation bias of eigenvalues of the covariance matrix due to finite amount of data. Section \ref{section:sampling} discusses uniform sampling of correlation stress tests and introduces an algorithm for creating such samples.
Section \ref{section:lax_stress} reflects on certain features of the changes in the covariance matrix defined by a Lax-pair evolution and show that they would be  less suitable as a model for correlation stress tests.
Section \ref{section:fullexample} illustrates the main concepts using a stylised example of bonds, equities and commodities.  The results are summarised in Section \ref{section:summary}.

\section{The Fisher--Rao manifold}
The Fisher--Rao metric for a family of probability distributions $p(X;\theta_i)$ on an $n$-dimensional space of random variables $X$ parametrised by $\theta_i$, $i=1,\dots, k$ is given by \cite{Rao_1945,Miyamoto_2024}:
\begin{equation}
    g_{ij}(\theta) = - \int _{\mathbb {R}^n}{\frac {\partial^2 \log p(X ;\theta ) }{\partial \theta_i \partial \theta_j}} p(X;\theta )dX.
\end{equation}
It is, under fairly general assumptions, the unique metric that does not depend on either the selection of the  underlying variables and or on the representation of the parameters of the distributions. The corresponding length is known as the Rao length. Its line element, with a slight abuse of notation, can be expressed as 
\begin{equation} \label{lab:line_element}
    d^2 = \int _{\mathbb {R}^n} \left(d \log{p(X;\theta)} \right)^2 p(X;\theta) dX.
\end{equation}
More specifically, consider the family of multivariate Gaussian distributions with vector of means $\mu$ and covariance $\Sigma$, and assume that both $\mu$ and $\Sigma$ are differentiable functions of some arbitrary parameters $\theta_i$, $i=1,\dots, k$ . The Fisher--Rao metric   reduces to \cite{atkinson_mitchell_1981,bib:fisher_rao_normal}:
\begin{equation}
\begin{split} \label{lab:fisher_rao_metric}
    g_{ij}(\theta)  = & \frac{\partial \mu^T}{\partial \theta_i} \Sigma^{-1} \frac{\partial \mu}{\partial \theta_j}  +  \\
     & 
    \frac{1}{2} \Tr \left( 
    \Sigma^{-1} 
    \frac{\partial \Sigma}{\partial \theta_i}
        \Sigma^{-1} 
      \frac{\partial \Sigma}{\partial \theta_j}
    \right) .
\end{split}
\end{equation}
In what follows we will consider the case of $\partial\mu/\partial \theta_i=0$ for all $i$, i.e. the submanifold of multivariate normal distributions with constant means.
Then, the Fisher--Rao distance between two distributions with covariances $\Sigma_2$ and $\Sigma_1$ in the $n(n+1)/2$-dimensional totally geodesic submanifold of Gaussian distributions with constant drifts is \cite{atkinson_mitchell_1981}:  
	\begin{equation} \label{lab:geodesic_distance}
	d^2_\mu( \Sigma_1, \Sigma_2)= \frac{1}{2} \sum_{i=1}^n \left[\log{(\lambda_i)}\right]^2,
	\end{equation}
	where $\lambda_i$ are the eigenvalues of $\Sigma_1^{-1} \Sigma_2$.
The unique curve of the shortest Rao length, the geodesic $\gamma(t)$, that connects two sufficiently close matrices $\Sigma_1$ at $t=0$ to $\Sigma_2$ at $t=1$ lies entirely within the multivariate  Gaussian submanifold with constant means of the (full) Fisher--Rao manifold. It can be explicitly constructed as \cite{moakher_geodesic,bib:fisher_rao_normal}:
\begin{equation} \label{lab:fisher_rao_geodesic}
    \gamma(t) = \Sigma_1^\frac{1}{2} \exp \left[ t \log \left( \Sigma_1^{-\frac{1}{2}} \Sigma_2 \Sigma_1^{-\frac{1}{2}} \right) \right] \Sigma_1^\frac{1}{2},
\end{equation}
where $\Sigma_1^\frac{1}{2}$ denotes the unique symmetric positive-definite matrix such that $ (\Sigma_1^\frac{1}{2})^2 = \Sigma_1$.

For a recent review of the geometry of probability distributions and some if its applications, see Ref. \cite{Nielsen_2024} for the slightly more general case of elliptically symmetric distributions, or Ref.  \cite{bouchard2023fisherraogeometrycesdistributions} for an overview of applications for covariance matrix estimation and classification using distance between statistical models and samples.

\section{Covariance and correlation stress tests} \label{section:correlation_stress_tests}
A covariance between risk factors can be thought of as a symmetric positive definite bilinear form on the space of risk factors. A change of basis in the space of risk factors will result in a change of entries in the matrix  representing the covariance form, but this will not change the covariance form itself. 

However, a change of entries of the covariance matrix which does not result from a change basis  will correspond to a different covariance form. Any change to the covariance form can be thought of as a covariance stress test. We will assume that the  stress tests depend on a parameter $t$ that can be interpreted as corresponding to the size of the stress.

We will define a \textit{covariance stress test} as a one-parameter family of functions that map the space of symmetric positive definite bilinear forms in $\mathbb{R}^n$ into itself. A \textit{correlation stress test} will be a covariance stress that does not change the  determinant of the covariance matrix along each path within the family. Thus, we define a correlation  stress test as a covariance stress that preserves the generalized variance of Wilks \cite{wilks_1932}.

For convenience, $\Sigma(0)$ will refer to the unstressed, or original, covariance matrix, while $\Sigma(t)$ will be the stressed covariance matrix.

One could object to calling a so-defined stress a \textit{correlation} stress test, and require instead that all the diagonal elements of the covariance matrix remain unchanged. The diagonal elements of the covariance matrix are defined with respect to a specific basis, so this would not be a basis-independent definition. Moreover, for any two symmetric positive definite matrices with the same determinant, there exists a basis \cite{biedrzycki_2024} in which they have the same diagonal elements.
Thus,  a covariance stress test which does not change the determinant of the covariance matrix can always be interpreted as keeping, \textit{in some basis}, all of the diagonal elements the same and  just changing the correlations, which is consistent with the intuitive understanding of stressing correlations. In conclusion, requiring no change of the determinant results in  defining a correlation stress test as a covariance stress test that, \textit{in some basis}, only  changes only the correlations between risk factors while keeping all variances the same. 

 A geometric interpretation provides some further guidance: the determinant of the covariance matrix is, for a multivariate normal distribution, the volume of an ellipsoid of constant probability density enveloping a given probability mass. The directions of the semi-axes of this ellipsoid represent the uncorrelated directions (principal components), while the length of a semi-axis represents the standard deviation of the principal component. A change in the covariance matrix that preserves the determinant would thus preserve the probability mass within it, or its generalised variance, while potentially changing the directions, and possibly also the lengths, of the semi-axes. This seems to be the closest analogue to the simple intuitive notion that  can be formulated without reference to a specific choice of variables used to represent the probability distribution. 

Equivalently, we could define a correlation stress test as a change in the covariance matrix that does not change the entropy 
$S = \frac{n}{2} + \frac{n}{2} \ln{2 \pi} + \frac{1}{2} \ln{\det \Sigma} $ 
of the Gaussian distribution with covariance  $\Sigma$. As our correlation stress tests preserve entropy, they could also be called adiabatic stress tests in analogy with thermodynamics, where  the adiabatic  processes do not change entropy. This definition is somewhat more general than the one that refers explicitly to the covariance matrix, since entropy can be defined for any probability distribution,  $S = -E\left(\log{p}\right) $.

An isotropic stress  which scales all the elements of the covariance matrix by the same positive number obviously does not change any correlations. Given two covariance matrices, one can always rescale one of them so that its determinant is equal to the determinant of the other. Thus, any covariance stress can be expressed as a correlation stress defined above and scaling of all entries of the covariance matrix by the same amount.


The following stylised example is perhaps useful in providing some further intuition. The example shows that what is clearly a correlation stress test in one basis ends up changing the variances in another, equally natural, basis. Consider a portfolio that consists of  equities and bonds. Assume the volatility of equities is 12\% and the volatility of price of bonds is 6\%,  and the bond--equity return correlation is 0. It is natural to consider returns of equities and bonds as the risk factors, which corresponds to the bond--equity set of market factors. But or course this is  not a unique set of possible market factors for this portfolio. Now, consider another set of variables that could be used:  a combination of returns of a balanced portfolio of 60\% of equities and 40\% of bonds, and the equity-bond spread, i.e., a factor which is 100\% equity returns less 100\% bond returns. Let's call it a balanced--spread set of market factors. As described, this new set of market factors has been derived from the original set by a change of variables.
The original covariance matrix in the bond--equity basis is:
\begin{equation}
    \Sigma =0.0001 \begin{pmatrix}
    144 &  0 \\
    0 &  36\\
    \end{pmatrix}.
\end{equation}
In the balanced--spread basis the covariance matrix is
\begin{equation}
    \Sigma =0.0001 \begin{pmatrix}
    57.6 &  72 \\
    72 &  180 \\
    \end{pmatrix}.
\end{equation}
Now suppose the correlation between bonds and equities is stressed to $0.1$. 
The respective covariance matrices become
\begin{equation}
    \Sigma =0.0001 \begin{pmatrix}
    144 &  7.2 \\
    7.2 &  36\\
    \end{pmatrix}
\end{equation}
in the bond--equity basis, and
\begin{equation}
    \Sigma =0.0001 \begin{pmatrix}
    61.056 &  70.56 \\
    70.56 &  165.60\\
    \end{pmatrix}
\end{equation}
in the balanced--spread basis.
What appears to be a correlation stress test in the bond--equity basis does not seem to appear as a correlation  stress in the balanced--spread basis since the  diagonal elements in the latter basis have  changed. Additionally, the determinant of the covariance matrix has changed from $5.184\times 10^{-5}$ to $5.132\times 10^{-5}$ (to three decimal places), and thus the probability mass enveloped by the ellipsoid of a constant probability density, as discussed above, has also changed. This example shows that if one aims for correlation stress tests that are more universal and do not depend on such arbitrary choices of basis of risk factors as in the above example, the intuitive approach needs to be refined.

As a side comment, one may observe that given two symmetric positive definite matrices $\Sigma_1$ and $\Sigma_2$ there  always exists a basis such that the entries of $\Sigma_2$ in this new basis are the same as the entries of $\Sigma_1$ in the original basis. This can be shown by applying the change of basis $V = \Sigma_2^{-\frac{1}{2}} \Sigma_1^\frac{1}{2} $, since
\begin{equation}
V^T \Sigma_2 V  = \left( \Sigma_2^{-\frac{1}{2}} \Sigma_1^\frac{1}{2}\right)^T 
\Sigma_2 
\left( \Sigma_2^{-\frac{1}{2}} \Sigma_1^\frac{1}{2}\right) = \Sigma_1.
\end{equation}
Thus, any stress test of the covariance matrix, naturally interpreted as an ``active'' change of the covariance matrix, can always be represented as a ``passive'' change of the basis in the vector space of the market risk factors. Such a change in general involves stretching the basis vectors,  as well as rotating them. It cannot be represented by a purely orthogonal transformation and isotropic scaling except for the trivial case in which the covariance matrices commute, $\left[ \Sigma_1, \Sigma_2 \right] = 0$. 
As a result, at a fixed point in the Fisher--Rao manifold, the space of infinitesimal covariance stress tests is isomorphic with the group $GL(n,\mathbb{R})$, while the space of correlation stress tests is isomorphic with the group $SL(n, \mathbb{R})$. 
In a sense, any covariance stress test can be thought of as ``some market factors just start behaving like some other market factors'', for example bonds beginning to move more like equities.
\section{Correlation stress tests as geodesics in the Fisher--Rao manifold} \label{section:geodesics}
Since
\begin{equation}
 \det \left( \exp{A} \right) = \exp \left( \Tr A \right)
 \end{equation}
 for any square matrix $A$, and 
 \begin{equation}
 \det B = \exp{\left( \Tr \log{B}\right) }
 \end{equation}
 for an invertible positive-definite matrix $B$, it follows from  Eq. \ref{lab:fisher_rao_geodesic} that the determinant of a covariance matrix along a geodesic path  is a simple function of the parameter $t$,
 \begin{align*}
\det \gamma(t)  = &  \det \left[ \Sigma_1^\frac{1}{2} \exp \left[ t \log \left( \Sigma_1^{-\frac{1}{2}} \Sigma_2 \Sigma_1^{-\frac{1}{2}} \right) \right] \Sigma_1^\frac{1}{2} \right] \\
=  & \det \left[ \exp \left[ t \log \left( \Sigma_1^{-\frac{1}{2}} \Sigma_2 \Sigma_1^{-\frac{1}{2}} \right) \right]  \right]  \det \Sigma_1   \\
= & 
\exp{\left[ t \Tr \left(  \log \left( \Sigma_1^{-\frac{1}{2}} \Sigma_2 \Sigma_1^{-\frac{1}{2}} \right)    \right) \right]} \det  \Sigma_1  \\
= &   
\left( \exp{\left[  \Tr \left(  \log \left( \Sigma_1^{-\frac{1}{2}} \Sigma_2 \Sigma_1^{-\frac{1}{2}} \right)    \right) \right]} \right)^t \det \Sigma_1\\
=& 
\left( \det \left( \Sigma_1^{-\frac{1}{2}} \Sigma_2 \Sigma_1^{-\frac{1}{2}} \right)  \right) ^ t \det  \Sigma_1   \\
= & \det \left( \Sigma_1^{-1}\right)^t \left( \det \left( \Sigma_2  \right) \right)^t \det \Sigma_1    \\
= & \left( \det \Sigma_1  \right)^{1-t} 
\left( \det \Sigma_2  \right)^t.
 \end{align*}
 In particular, a geodesic joining two covariance matrices whose determinant is the same consists of matrices of the same determinant. 
 \begin{equation}
 \det \gamma(t)  = \det \Sigma_1 = \det \Sigma_2.
 \end{equation}
 Select a point $\Sigma_1$ in the Fisher--Rao manifold of Gaussian distributions corresponding to all covariance matrices. The sub-manifold passing through $\Sigma_1$ that consists of all matrices $\Sigma$ with the same determinant, $\det \Sigma = \det \Sigma_1$, is thus spanned by the geodesics emanating from $\Sigma_1$ that lie entirely within that submanifold, and so define an exponential map within that submanifold. As a result of the geodesics always staying within this submanifold, the extrinsic curvature of the submanifold of covariance matrices with constant determinant, as embedded in the Fisher--Rao manifold,  vanishes  \cite{jurek_2024}.  
 
In the next section, we will show that the exponential map can be extended to arbitrarily long geodesics,  and thus defines arbitrarily large correlation stress tests.

\section{Parametrisations of the sub-manifold of the correlation stress tests} \label{section:parametrisation}
Skovgaard's \cite{skovgaard_1984} Theorem 6.7 shows how a geodesic can be constructed by diagonalisation with respect to the original covariance matrix, but this original formulation may not have an apparent intuitive link to the correlation stress tests. It is perhaps more natural to consider, as in Ref. \cite{moakher_geodesic}, a one-parameter family of covariance matrices  $\Sigma(t)$  which are a result of stressing   $\Sigma = \Sigma(0)$ to $\Sigma(t)$ 
\begin{equation}  \label{lab:stress_test_path}
 \Sigma(t) = \Sigma_X(t) = \Sigma^\frac{1}{2}\exp{(X t)} \Sigma^\frac{1}{2},
\end{equation}
where the subscript in $\Sigma_X(t)$ indicates the explicit dependence of this  family on  a symmetric traceless matrix $X$, and the parameter $t$ corresponds to the amount of stress.  Thus, $\Sigma(t)$ is symmetric and positive definite. Also, since
\begin{equation} 
\det(\exp(Xt)) = \exp( \Tr (Xt)) = 1,
\end{equation}
$\det\Sigma(t) = \det\Sigma(0)$, so Eq. \ref{lab:stress_test_path} indeed defines a correlation stress and a geodesic in the Fisher--Rao manifold.

This can be explicitly confirmed substituting Eq. \ref{lab:stress_test_path}  into Eq. \ref{lab:fisher_rao_geodesic} and setting
 $\Sigma_1=\Sigma(0)$ and  $\Sigma_2=\Sigma(1)$ to show \cite{moakher_geodesic} that
\begin{flalign}  
\gamma(t) & =  \Sigma_1^\frac{1}{2} \nonumber
\exp{\left[ t  \log{  \left( \Sigma_1^{-\frac{1}{2}} \Sigma_1^{\frac{1}{2}} \exp{(X)} \Sigma_1^{\frac{1}{2}}  \Sigma_1^{-\frac{1}{2}} \right)  }\right]} 
\Sigma^\frac{1}{2} \\ \nonumber
          & =
                \Sigma^\frac{1}{2} \exp \left[ t \log  
                \exp{(X)} 
                \right] \Sigma^\frac{1}{2}  \\  \nonumber
          & = \Sigma^\frac{1}{2}\exp{(X t)} \Sigma^\frac{1}{2} \\ 
          & =  \Sigma(t). 
 \end{flalign}
 Thus Eq. \ref{lab:stress_test_path} is indeed a geodesic parametrised by $t$. Alternatively, this can be confirmed by applying Theorem 6.1 of Ref. \cite{skovgaard_1984} to $\Sigma(t)$ defined in Eq. \ref{lab:stress_test_path}. For a choice of basis in the space of  matrices $X$ that span the space of all traceless symmetric matrices of order $n$, Eq. \ref{lab:stress_test_path} defines explicitly the exponential map in the sub-manifold of correlation stress tests.

Eq. \ref{lab:stress_test_path} is formally equivalent to Eq. \ref{lab:fisher_rao_geodesic}, but rather than depending explicitly on the end points of the geodesic, it is explicitly parametrised by the starting point and an arbitrary trace-free symmetric matrix which corresponds to the tangent space to the Fisher--Rao manifold at the point $\Sigma(0)$. In addition to the path defined in Eq. \ref{lab:stress_test_path}, consider another path parametrised by a trace-free symmetric matrix $Y$,
\begin{equation}  
 \Sigma_Y(t) = \Sigma^\frac{1}{2}\exp{(Y t)} \Sigma^\frac{1}{2}.
\end{equation}
The  scalar product at $\Sigma(0)$ defined  by Eq. \ref{lab:fisher_rao_metric} reduces to 
\begin{align} 
    g(X,Y)  = &   \frac{1}{2} \Tr \left( 
    \Sigma^{-1} 
    \frac{\partial \Sigma_X}{\partial t}
        \Sigma^{-1} 
      \frac{\partial \Sigma_Y}{\partial t}
    \right) \nonumber \\
    = & \frac{1}{2} \Tr (X Y), \label{equation:tangent_product}
    \end{align}
i.e. the Frobenius norm on the space of the traceless matrices $X$ that define tangent space at $\Sigma$. 
Equation \ref{lab:stress_test_path} can of course be also written as 
\begin{equation}
    \Sigma(t) = \Gamma^T(t) \Gamma(t)
\end{equation}
with $\Gamma(t) = \exp{ \left( \frac{1}{2} Xt \right) } \Sigma^\frac{1}{2}$. 

As $\Sigma^\frac{1}{2}$ is the matrix of change of basis into one in which the covariance matrix is the identity matrix, one can interpret the above parametrisation as implicitly referring to changing the correlations away from the identity matrix, when expressed in the basis of the normalised eigenvectors of $\Sigma$. 

Various choices of $X$ can be made. A specific selection of the entries of $X$ will define a stress test with respect to a specific choice of basis in the space of the random variables. A practitioner may have a portfolio of special concern, whose exposures may be known in a defined basis. Or perhaps, considering a narrative of what may affect the markets, they may be concerned with the behaviour of a number of portfolios, some of which may be affected in different ways. In the following sections, we will consider a measure of size of a stress test --- its plausibility, which allows one to have an a priori view on what size stress tests need to be considered, without specifying the portfolio and its market factors themselves.

 \section{Plausibility of correlation stress tests} \label{section:plausibility}

The line element defined by Eq. \ref{lab:line_element} is the expected value of the change of the log-probability corresponding to an infinitesimal change of the distribution. Intuitively, the more a distribution is changed, the less plausible the new distribution may be. The geodesic distance $d$  between two distributions is a natural measure of how much these two distributions differ. It may be thought of as the integral of the infinitesimal differences between their logarithms, hence the exponential function of that distance can  be interpreted as a measure of the  \textit{plausibility} of a stress test that morphs a distribution $p_1(X)$ into another distribution $p_2(X)$, 
\begin{equation} \label{lab:plausibility}
    P\left( p_1  \rightarrow p_2 \right) = \exp{(-d)}.
\end{equation}
The plausibility will range between zero and one, the former corresponding to an infinitely large stress, and the latter to no stress at all.  

For multivariate normal distributions, the plausibility of a stress test that morphs  $\Sigma_1$ into $\Sigma_2$ is
\begin{equation} 
    P\left( \Sigma_1  \rightarrow \Sigma_2 \right) = \exp{(-d)}.
\end{equation}
For a specific correlation stress test path defined in Eq. \ref{lab:stress_test_path}, this distance, Eq. \ref{lab:geodesic_distance}, can be calculated using the eigenvalues $\lambda_i$ of 
$\Sigma^{-1/2}\exp{(X t)} \Sigma^{1/2}$. Let $x_i$ be an eigenvalue corresponding to an eigenvector $v_i$ of the above matrix, so the eigenvalue equation is
\begin{equation}
\Sigma^{-1/2}\exp{(X t)} \Sigma^{1/2} v_i =x_i v_i.
\end{equation}
Multiplying both sides by $\Sigma^{1/2}$ and setting $\Sigma^{1/2} v_i = u_i$ shows that the eigenvalues $x_i$ satisfy 
\begin{equation}
    \exp{(X t)} u_i = x_i u_i.
\end{equation}
Thus $x_i$ is an eigenvalue of $\exp(Xt)$, or $\log(x_i)$ is an eigenvalue of $X$ multiplied by $t$.  Then Eq. \ref{lab:geodesic_distance} implies 
\begin{equation} \label{lab:correlation_stress_distance}
    d^2 \left( \Sigma(0) , \Sigma(t) \right) = \frac{t^2}{2} \sum_{i=1}^n  x_i^2 = \frac{t^2}{2} \Tr{\left( X^2\right)}
\end{equation}
with $x_i$ being the eigenvalues of $X$. Since the values of $d^2$ can be arbitrarily large for some   $t \in \mathbb {R}$,  the submanifold of correlation stress tests is geodesically complete.

Using Eq. \ref{lab:plausibility},  the plausibility of a correlation stress test defined by stressing $\Sigma(0)$ to $\Sigma(t)$ as defined by Eq. \ref{lab:stress_test_path} is
\begin{equation} \label{lab:plausibility1}
    P(X, t) = \exp{\left( - \sqrt{ \frac{t^2}{2} \sum_{i=1}^n  x_i^2 }  \right) }.
\end{equation}

More plausible stress tests would perhaps be of higher concern to a risk manager. Specifying a plausibility threshold level, between zero and one, is akin to explicitly specifying a level of risk tolerance to stress tests. One can also use Eqs. \ref{lab:plausibility} and \ref{lab:plausibility1} to assess relative plausibility of one stress test compared to another. For example, a stress test that is tenfold, or hundredfold,  as plausible as another stress test, may perhaps be considered as more relevant. 

Note that the plausibility of a correlation stress test does not depend on the initial covariance matrix as $\Sigma(0)$ does not enter Eq. \ref{lab:plausibility1}. Thus, the plausibility of a correlation stress test is in a sense universal: different covariance matrices can be subjected to the same correlation stress test of a given plausibility, and stress tests can be ranked according to their plausibility irrespective of the initial unstressed covariance matrix.

\section{Stressing a single pair of correlations} \label{section:single_pair_stress}

If one selects a specific basis of the random variables, one can think of stressing a correlation of two variables indexed by $a$ and $b$, $a\neq b$,  as defined by  by Eq. \ref{lab:stress_test_path} with $X_{ij} = \delta^a_{(i}\delta^b_{j)}$ where the brackets 
denote symmetrisation with respect to the respective indices. Without loss of generality, one can assume that $a=1$ and $b=2$, and represent $X$ in a block-diagonal form  where the entries with $X_{i,j} =0$ for $i,j>2$ except for a block where $i \leq 2$ and $j \leq 2 $. Then
\begin{equation}
X=
\begin{pmatrix}
    0,&  1, & 0, & \cdots &,& 0 \\
    1, & 0, & 0, & \cdots &,& 0 \\
    0, & 0, & 0, & \cdots &,& 0 \\
     &  & \cdots &  & &  \\
    0, & 0, & 0, & \cdots &,& 0 \\
\end{pmatrix},
\end{equation}
and
\begin{equation}
X^2=
\begin{pmatrix}
    1,&  0, & 0, & \cdots &,& 0 \\
    0, & 1, & 0, & \cdots &,& 0 \\
    0, & 0, & 0, & \cdots &,& 0 \\
     &  & \cdots &  & &  \\
    0, & 0, & 0, & \cdots &,& 0 \\
\end{pmatrix}.
\end{equation}
As a consequence, $\exp{( t X)} $ is also block-diagonal with the same structure,
\begin{equation}
e^{ t X }=
\begin{pmatrix}
    \cosh{t}, &  \sinh{t},  & 0, & \cdots &,& 0 \\
    \sinh{t}, &  \cosh{t}  & 0, & \cdots &,& 0 \\
    0, & 0, & 1, & \cdots &,& 0 \\
     &  & \cdots &  & &  \\
    0, & 0, & 0, & \cdots &,& 1 \\
\end{pmatrix}.
\end{equation}
If, in addition, in that specific basis the covariance matrix $\Sigma$ is diagonal, $\Sigma = \diag(\sigma_i^2)$, then the stressed covariance matrix $\Sigma(t)$ is itself block-diagonal 
\begin{equation} \nonumber
\Sigma(t) =
\begin{pmatrix}
     \sigma_1^2 \cosh{t}, &   \sigma_1 \sigma_2  \sinh{t},  & 0, & \cdots &,& 0 \\
     \sigma_1 \sigma_2  \sinh{t}, &  \sigma_2^2  \cosh{t}  & 0, & \cdots &,& 0 \\
    0, & 0, & \sigma_3^2, & \cdots &,& 0 \\
     &  & \cdots &  & &  \\
    0, & 0, & 0, & \cdots &,& \sigma_n^2 \\
\end{pmatrix}.
\end{equation}
The entries $\sigma_j^2$ for $j>2$ are unaffected by the stress test. While the infinitesimal stress test $X$ only impacts the two entries corresponding to the correlations of the two variables, in a finite stress test, all covariances in the top-left block, including the diagonal elements, are affected by this correlation stress, although of course the determinant of the covariance matrix does not change. If the stress test is with respect to a basis in which the covariance matrix is diagonal, only the block corresponding to the variables whose correlation is stressed is affected, while the remaining covariances stay the same, which one would intuitively expect. This is not the case if $\Sigma^\frac{1}{2}$ is not diagonal, as all correlations are effectively mixed and interrelated in such stress tests.

Note that in the above stress test, the diagonal elements are unchanged up to first order of $t$,  thus the variances and standard deviations are not affected up to order $O(t)$ by the infinitesimal version of this stress test. 

\section{Stressing along the diagonal} \label{section:diagonal_stress}
Of course a traceless matrix $X$ needs not have all zero diagonal elements. Without loss of generality, a convenient choice of a basis of the stress tests within the diagonal can be parametrised by setting $X_{11}=1$ and $X_{22}=-1$, with all other entries being $0$:
\begin{equation}
X=
\begin{pmatrix}
    1,&  0, & 0, & \cdots &,& 0 \\
    0, & -1, & 0, & \cdots &,& 0 \\
    0, & 0, & 0, & \cdots &,& 0 \\
     &  & \cdots &  & &  \\
    0, & 0, & 0, & \cdots &,& 0 \\
\end{pmatrix}.
\end{equation}
Thus
\begin{equation}
X^2=
\begin{pmatrix}
    1,&  0, & 0, & \cdots &,& 0 \\
    0, & 1, & 0, & \cdots &,& 0 \\
    0, & 0, & 0, & \cdots &,& 0 \\
     &  & \cdots &  & &  \\
    0, & 0, & 0, & \cdots &,& 0 \\
\end{pmatrix}.
\end{equation}

As a consequence, $\exp{( t X)} $ is also block-diagonal,
\begin{equation}
e^{ t X }=
\begin{pmatrix}
    \exp{(t)} , &  0,  & 0, & \cdots &,& 0 \\
    0, &  \exp{(-t)}, & 0, & \cdots &,& 0 \\
    0, & 0, & 1, & \cdots &,& 0 \\
     &  & \cdots &  & &  \\
    0, & 0, & 0, & \cdots &,& 1 \\
\end{pmatrix}.
\end{equation}

If such a stress is applied to a covariance  matrix $\Sigma$ that is diagonal, $\Sigma = \diag(\sigma_i^2)$, the stressed covariance matrix $\Sigma(t)$ is itself block-diagonal

\begin{equation} \nonumber
\Sigma(t) =
\begin{pmatrix}
     \sigma_1^2 \exp{(t)}, &   0,  & 0, & \cdots &,& 0 \\
     0, &  \sigma_2^2  \exp{(-t)}  & 0, & \cdots &,& 0 \\
    0, & 0, & \sigma_3^2, & \cdots &,& 0 \\
     &  & \cdots &  & &  \\
    0, & 0, & 0, & \cdots &,& \sigma_n^2 \\
\end{pmatrix}.
\end{equation}

The stress test described above may not be intuitively perceived as a correlation stress. However, together with the stress tests described in the previous section, the matrices $X$ may be used to form a basis of the tangent space at any covariance matrix $\Sigma(0)$. Thus, the examples in these two sections describe a basis of the most general correlation stress tests.

\section{Stressing all correlations of a single variable } \label{section:all_correlations_of_single_variable}
In a selected basis of the random variables, one can think of stressing all correlations of a single  factor  with all other factors by an equal amount, with all other correlations unchanged. Without loss of generality, assume that this variable is the first variable, so that $X$  can be represented as:
\begin{equation}
X=
\begin{pmatrix}
    0,&  1, & 1, & \cdots &,& 1 \\
    1, & 0, & 0, & \cdots &,& 0 \\
    1, & 0, & 0, & \cdots &,& 0 \\
     &  & \cdots &  & &  \\
    1, & 0, & 0, & \cdots &,& 0 \\
\end{pmatrix},
\end{equation}
so 
\begin{equation}
X^2=  
\begin{pmatrix}
    n-1,&  0, & 0, & \cdots &,& 0 \\
    0, &1, & 1, & \cdots &,& 1 \\
    0, & 1, & 1, & \cdots &,& 1 \\
     &  & \cdots &  & &  \\
    0, & 1, & 1, & \cdots &,& 1 \\
\end{pmatrix}.
\end{equation}
Let $Z = X^2$. Then
\begin{equation}
    X^k = \begin{dcases*}
    (n-1)^{\frac{k}{2}-1} Z & when  $k$ is even \\
    (n-1)^\frac{k-1}{2} X   & when  $k$ is odd.
    \end{dcases*}
\end{equation} 
The power series expansion of 
$\exp{(  t X )}$ can then be expressed as a linear combination of X and Z,
\begin{align}
    \exp{( t X  )} = & I + \frac{1}{\sqrt{n-1}} X \sinh{(\sqrt{n-1}t)} \\ 
     & +   \frac{1}{n-1} Z \left( \cosh{(\sqrt{n-1}t)} -1 \right)
\end{align}
where $I$ is the $n \times n$ identity matrix.

\section{Equally stressing all correlations } \label{section:bias_of_eigenvalues}
In the previous sections we have discussed several stress tests resulting from specific choices of the matrix $X$ generating the paths given by Eq. \ref{lab:stress_test_path}. In this section we will consider the case of a  hollow matrix $X$ whose non-diagonal entries are all equal, and will argue that this example can be considered as a stress test corresponding to a change of the sample size of the data. 

Lawley \cite{lawley1956} showed that, for a sample of size $m$, the estimates $\hat{\lambda}_r$ of the eigenvalues of the covariance matrix of $p$ random variables are biased relative to the eigenvalues $\lambda_r$ of the population covariance,
\begin{align}
\label{lab:lawleymap} \nonumber
\hat{\lambda}_r= & \lambda_r\left\{ 1 + \frac{1}{m} \sum^p_{i=1; i\ne r} \frac{\lambda_i}{\lambda_r-\lambda_i} \right\} \\
& + \mathcal{O}\left(m^{-2}\right).
\end{align}
Thus, the trace of the estimate is equal to the trace of the population covariance matrix, up to order of $O\left(m^{-2}\right)$.

Consider an infinitesimal version of a specific stress test of Eq. \ref{lab:stress_test_path} where  all off-diagonal elements of $X$ are equal.
This can be shown as follows: set 
\begin{equation} \label{lab:all_equal}
X=
\begin{pmatrix}
    0,&  1, & 1, & \cdots &,& 1 \\
    1, & 0, & 1, & \cdots &,& 1 \\
    1, & 1, & 0, & \cdots &,& 1 \\
     &  & \cdots &  & &  \\
    1, & 1, & 1, & \cdots &,& 0\\
\end{pmatrix},
\end{equation}
in the basis of the eigenvectors of $\Sigma(0)$, so $\Sigma(0) = \diag(\sigma_i^2)$. This corresponds to setting $l_r = \sigma_r^2$ in Eq. \ref{lab:lawleymap}.
Consider the first two terms in the power series expansion in $t$ of Eq. \ref{lab:stress_test_path}: 
\begin{equation} \label{lab:linear_stress_path}
    \Sigma(t) = \Sigma^\frac{1}{2}  \left( I + Xt \right) \Sigma^\frac{1}{2}.
\end{equation}
By using the power series expansion in $t$ of the eigenvalues of $\Sigma(t)$ we will show that they have the same functional form as Eq. \ref{lab:lawleymap}.

Ref. \cite{tta_blog_eigenval_2017} discusses the well-known feature of the impact of perturbations on the spectrum of symmetric positive-definite matrices:  the  eigenvalues $\lambda_i$ of a one-parameter family of matrices $A(t)$ behave as if they were subject to a mutually repulsive force:
\begin{equation} \label{lab:firstOrderCorrection}
\dot{\lambda}_i=u_i^T \dot{A} u_i
\end{equation}
and
\begin{equation}  \label{lab:secondOrderCorrection}
\ddot{\lambda}_i=u_i^T \ddot{A} u_i + 2 \sum_{j \ne i}{ \frac{| u_j^T \dot{A} u_i | ^2}{\lambda_i - \lambda_j}},
\end{equation}
where the dot denotes differentiation with respect to the parameter $t$, two dots denote the second derivative, and $u_i$ are the normalised eigenvectors of $A(0)$.

Set $ A(t) = \Sigma(t)$, and let $u_i$ be the normalised eigenvectors of $\Sigma(0)$. The first derivative of $\Sigma(t)$ given by Eq. \ref{lab:linear_stress_path} is
\begin{equation}
\dot{\Sigma}(0) = \Sigma^\frac{1}{2} X \Sigma^\frac{1}{2},
\end{equation}
where $\Sigma^\frac{1}{2} = \diag(\sigma_i)$. All diagonal entries of the matrix $X$, and hence of $\dot{\Sigma}(0)$,  are  zero, all off-diagonal entries of  $\dot{\Sigma}(0)$ are $\sigma_j\sigma_k$, and all components of the vectors $u_i$ are zero except for the $i$-th component which is $1$. As a result   $u_i^T\dot{\Sigma}(0) u_i= 0$, while   $u_j^T\dot{\Sigma}(0) u_k = \sigma_j \sigma_k $ for $j \neq k$.
Separately, the second derivative of $\Sigma(t)$ given by Eq. \ref{lab:linear_stress_path} is zero. 
These observations have the following two implications.

First,    as a result of Eq. \ref{lab:firstOrderCorrection}, the first derivative of the eigenvalues of $\Sigma(t)$ defined by Eq. \ref{lab:linear_stress_path}  vanish at $t=0$. Hence, there is no first-order term in the power-series expansion of the eigenvalues of $\Sigma(t)$ around $t=0$.

Second,  since  $\ddot{\Sigma}(t)$ of Eq. \ref{lab:linear_stress_path} vanishes, Eq. \ref{lab:secondOrderCorrection} shows that the second derivative of the eigenvalues of $\Sigma(t)$ defined by Eq. \ref{lab:linear_stress_path} is
\begin{equation}
    \ddot{\lambda}_i=2 \sum_{j \ne i}\frac{ \sigma_i^2\sigma_j^2}{\lambda_i - \lambda_j} = 
    2 \sum_{j \ne i}\frac{ \lambda_i\lambda_j}{\lambda_i - \lambda_j}.
\end{equation}
As a result, the first two terms in the power series expansion around $t=0$ of the eigenvalues of $\Sigma(t)$ have  the same functional form  as Eq. \ref{lab:lawleymap}, with $s=\sqrt{t}$.

Interestingly, the original article of Lawley \cite{lawley1956} considered the bias of eigenvalues due to finite amount of data. One can interpret the above result as implying that increasing (or decreasing) the amount of data may result in what is effectively  a correlation stress. Eq. \ref{lab:stress_test_path} with $X$ of Eq. \ref{lab:all_equal} allows such stress test to  be extended to an arbitrarily large size, or equivalently to arbitrarily low plausibility.

This type of correlation stress may also be useful for stress-testing of the valuation of derivatives with a relatively large number of similar underlying market factors, for example  a swap on volatility or variance of a basket or an option on a basket. For such baskets, the covariance matrix is of course likely to be non-diagonal, but the stress test defined by $X$ in Eq. \ref{lab:all_equal} can still be applied.

It may be worth noting that, for $X$ defined in Eq. \ref{lab:all_equal}, there is a relatively simple expression for $\exp{ \left( t X \right)}$. Adding the identity matrix $I$ to $X$ results in a matrix of ones $J$, which has a rank of one and  has only two eigenvalues, $n=\dim(X)$ of multiplicity of one, and zero as the remaining eigenvalues. By inspecting its eigenvectors, the matrix exponential of $tJ$ can be demonstrated to be 
\begin{equation}
    \exp{\left( t J \right) } = I + \frac{e^{nt}-1}{n}J, 
\end{equation}
so 
\begin{align}
    \exp{\left( t X \right)} =  \exp{\left(  Jt - I t \right)}  \\
    =  e^{-t} \left( I + \frac{e^{nt} -1}{n}  J \right).
\end{align}

\section{Uniform sampling of correlation stress tests} \label{section:sampling}
Eqs. \ref{lab:stress_test_path} and \ref{lab:plausibility1}  show how to define an arbitrary correlation stress test at a given plausibility level $P$. In practice, it may also be useful to be sample such stress tests in a way that does not rely on a specific  preference of one such a specific stress  over another. Thus, one would need to  ``uniformly'' sample correlation stress tests. Uniformity is closely related to measures and symmetries: a sample is uniformly distributed if it has the same number of points within sets of the same measure, and such a measure needs to be invariant under the symmetries of the problem. 

Here, the set of points is selected from the stressed covariances with the same plausibility, which are the points at a fixed geodesic distance from the unstressed matrix $\Sigma_0$.  These points form a normal sphere --- the submanifold of a given Rao distance from the unstressed matrix. The symmetry of this submanifold is directly related to the symmetry of the tangent space at $\Sigma_0$ which preserves the Frobenius scalar product, Eq. \ref{equation:tangent_product}. This can be shown as follows: select an arbitrary point on the normal sphere, and consider its pull back to the tangent space at $\Sigma_0$. Then, rotate the resulting tangent vector through a linear transformation which is orthogonal with respect to the Frobenius norm. Finally, push forward the so-rotated tangent vector to the normal sphere \footnote{This would normally be true only up to a certain value of the radius of the normal sphere. The Fisher--Rao manifold considered here has negative curvature however, so is true for any radius}. Thus, the group of symmetries of the normal sphere is the same as the group of symmetries of the scalar-product preserving symmetries of the tangent space. There is a unique, up to a multiplicative constant, measure that is invariant under this group,  the Haar measure. Moreover, both the group of transformations of the normal sphere and the group of orthogonal transformations of the tangent space act transitively, so the Haar measure of the group gives rise to the measure on the respective spaces on which it acts. Thus the measure on the normal sphere and the measure on the sphere in the tangent space are the same, up to a multiplicative constant, namely the standard measure on a sphere in the tangent space. 

The tangent space at $\Sigma_0$ consists of symmetric traceless matrices. The Frobenius norm can be expressed in the standard vector notation as 
\begin{equation}
    \Vec{X} \cdot \Vec{Y} = \Tr{ X Y},
\end{equation}
where $X$ and $Y$ are general $n \times n$ matrices viewed as vectors in $\mathrm{R}^{N}$, where $N=n^2$, and $n$ is number of risk factors. The unit sphere in this space is the set of points such that 
\begin{equation} \label{equation:unit_sphere}
     \Vec{X} \cdot \Vec{X} = \Tr{ X X} = 1.
\end{equation}
Eq. \ref{equation:unit_sphere} defines the unit sphere in the $N$-dimensional space of general $n$x$n$ matrices. Symmetric and trace-free matrices form a subspace thereof which is an orthogonal projection of $\mathrm{R}^{N}$ with respect to a set of mutually orthogonal vectors. Thus, the projection of the unit sphere in $\mathrm{R}^{N}$ is also a unit sphere of lower dimension, $n (n-1)/2 -1$. This can be shown as follows.

Let $E_{ab}$, with $a,b \in 1,\dots, n$ ,be matrices whose entries are all zero except for the entry, in $a$-th row and $b$-th column, which is $1$ and the entry in $b$-th row and $a$-th column, which is set to $-1$. Let $I$ be the identity matrix. Since
\begin{equation}
   \Vec{E}_{ab}\cdot \Vec{E}_{cd} = \Tr{ E_{ab} E_{cd}} = 0,
   \end{equation}
and 
 \begin{equation}
 \Vec{E}_{ab}\cdot \Vec{I} = \Tr{ E_{ab} I } = 0,
\end{equation}
the matrices $E_{ab}$ and $I$ span a set of mutually orthogonal vectors in vector space $R^N$ of general $n \times n$ matrices.

Suppose that a matrix $X$ satisfies
\begin{equation}
 \Vec{E}_{ab}\cdot \Vec{X} = 0,
\end{equation}
for any  $a,b \in 1,\dots, n$
and 
\begin{equation}
 \Vec{I} \cdot \Vec{X} = \Tr{ I X} =   0.
\end{equation}
Then $X$ is trivially symmetric and trace-free.
As the $N-1=n^2-1$ dimensional sphere consisting of all matrices $X$ such that 
\begin{equation}
   \Vec{X}\cdot \Vec{X} = \Tr{ X X } = 1
\end{equation}
is (spherically) symmetric, its projection onto the subspace orthogonal to all matrices $E_{ab}$ and $I$ will thus be also a sphere, in the natural metric induced from $R^N$ and thus provides the notion of uniformity for symmetric positive definite matrices.

The arguments presented above suggest the following algorithm for constructing a uniformly distributed set of $M$ correlation stress tests $\Sigma_s$ of an unstressed covariance matrix $\Sigma_0$:
\begin{enumerate}

    \item Select $N^2$ independently and normally distributed random numbers. Rescale them so that the sum of their squares is $1$. The rotational symmetry of the uncorrelated multivariate normal distribution implies that such numbers, when viewed as coordinates in $\mathrm{R}^N$, represent a matrix $X$ from a population that is uniformly distributed on a unit sphere $S(N-1)$. 
    \item For all $a, b \in 1,\dots, n$ with $a<b$, recursively project this matrix onto the sub-space orthogonal to each consecutive $E_{ab}$,
        \begin{equation}
            \Vec{X} \coloneqq \vec{X} - \frac{  \vec{X} \cdot \vec{E}_{ab}}{\vec{E}_{ab} \cdot \vec{E}_{ab} } E_{ab},
        \end{equation}
    where $\coloneqq$ denotes the assignment for the next recursion step.
  \item Project this matrix further onto the sub-space orthogonal to the unit matrix $I$:
    \begin{equation}
            \Vec{X} \coloneqq \vec{X} - \frac{  \vec{X} \cdot \vec{I}}{n } \Vec{I}.
        \end{equation}
  \item Finally, use Eq. \ref{lab:stress_test_path}
  \begin{equation}  
    \Sigma_s \coloneqq \Sigma_0^\frac{1}{2}\exp{(X t)} \Sigma_0^\frac{1}{2}
    \end{equation}
    to define the stressed covariance matrix which is a point on the normal sphere of radius defined by Eq. \ref{lab:correlation_stress_distance}
 \end{enumerate} 

Repeating the above steps $M$ times will result in a sample of uniformly distributed covariance matrices, each  corresponding to a correlation stress of the same plausibility, as defined in Eq. \ref{lab:plausibility1}.

An example of applying the above algorithm to a stylised case of portfolio of three assets is discussed in Section \ref{section:fullexample}.

\section{Lax pair stress tests} \label{section:lax_stress}
Instead of the approach to correlation stress tests discussed in Section \ref{section:correlation_stress_tests}, one could require a different, perhaps more stringent,  definition of a correlation stress test --- namely  that the eigenvalues of the covariance matrix do not change under such a stress test. 
In this section, we show that such an approach would be less suitable than that of Eq. \ref{lab:stress_test_path}.

Suppose, for simplicity, that the eigenvalues are distinct, so that the spectrum of $\Sigma(0)$ is non-degenerate.
The equality of eigenvalues along a path $\Sigma(t)$ would then mean that both $\Sigma(t)$ and $\Sigma(0)$, when diagonalised, are equal:
\begin{equation} \label{lab:equal_eigenvalues}
\left( V(t)\right)^T\Sigma(t) V(t) = \left( V(0) \right)^T \Sigma(0) V(0)
\end{equation}
for a non-singular orthogonal family of matrices $V(t)$.  One can take the derivative of Eq. \ref{lab:equal_eigenvalues} with respect to $t$, note that  $\left(V (t) \right) ^T = \left (V(t)\right) ^{-1}$  because $\Sigma(t)$ remains symmetric, and use
\begin{equation} 
    \frac{d V^T}{dt}  = \frac{d V^{-1}}{dt}  = -V^{-1} \frac{dV}{dt} V^{-1}.
\end{equation} 
This shows that $Y(t) = \frac{d V^T}{dt}  V^{-1}$ and  $\Sigma(t)$ are a Lax pair, 
\begin{equation} \label{lab:laxpair}
 \frac{d}{dt} \Sigma(t)= [Y(t), \Sigma(t)],
\end{equation}
where the square brackets denote the commutator of matrices, $[A, B] = AB - BA$.

Eq. \ref{lab:laxpair} is a first order differential equation that defines a path connecting $\Sigma(0)$ with $\Sigma(t)$. Instead of considering, as in Eq. \ref{lab:stress_test_path}, the geodesic paths, one could then perhaps consider stress tests defined by  the Lax pair, Eq. \ref{lab:laxpair}. 

We will now show that in most cases these two equations cannot describe the same path. We then will argue that this is a significant drawback of the Lax pair stress tests.

Suppose that the eigenvalues $\lambda_i(t)$ of $\Sigma(t)$ do not change in the neighbourhood of $t=0$. Select a basis of eigenvectors $u_i$ of $\Sigma(0)$. In this basis
$\Sigma= \Sigma(0) = \diag(\lambda_i) = \diag(\sigma_i^2)$ and
\begin{equation}
    \dot{\Sigma}(0) = \Sigma^\frac{1}{2} X \Sigma^\frac{1}{2}.
\end{equation}
Let $X_{ij}$ be the entries of the matrix $X$.

Eq. \ref{lab:firstOrderCorrection} implies that 
\begin{equation}
    \dot{\lambda}_i=u_i^T \dot{\Sigma(t)} u_i = u_i^T\Sigma^\frac{1}{2} X \Sigma^\frac{1}{2} u_i = X_{ii}^2 \sigma^2_i.
\end{equation}
Vanishing of $\dot{\lambda}_i$ then implies that the diagonal elements of $X$ vanish, $X_{ii}=0$.
Suppose that the spectrum of $\Sigma(0)$ is non-degenerate, so that no two eigenvalues $\lambda_i$ are equal. Select the eigenvalues in a decreasing order, so that $\lambda_1$ is the largest one.  Then Eq. \ref{lab:secondOrderCorrection} with 
\begin{equation}
    \ddot{\Sigma}(0) = \Sigma^\frac{1}{2} X^2 \Sigma^\frac{1}{2}
\end{equation}
implies that
\begin{align*}  
\ddot{\lambda}_i = & u_i^T \ddot{A} u_i + 2 \sum_{j \ne i}{ \frac{| u_j^T \dot{A} u_i | ^2} {\lambda_i - \lambda_j}}\\
=& \sigma_i^2\sum_k X_{ik}^2 + 2 \sum_{k \neq i} \frac{ X_{ik}^2 \sigma_k^2\sigma_i^2}{\lambda_i - \lambda_k} \\
=& \sigma_i^2 X_{ii}^2 + \sum_{k \neq i}\sigma_i^2  X_{ik}^2  \left(  1 + 2 \frac{  \sigma_k^2}{\lambda_i - \lambda_k} \right)\\
=& \sum_{k \neq i}\sigma_i^2  X_{ik}^2  \left(  1 + 2 \frac{  \sigma_k^2}{\lambda_i - \lambda_k} \right)
\end{align*}
since $X_{ii}=0$.

Start with the largest eigenvalue $\lambda_1$. Setting  $\ddot{\lambda}_1=0$  implies that  $X_{1k}=0$ for all $k=1,\dots, n$ since $\lambda_1-\lambda_k > 0$ for all $k$. Since $X$ is symmetric, this also implies $X_{k1}=0$, so $X$ must be a matrix with zeros in the first row and first column. The same argument can then be repeated to show that $X_{2k}=0$ for all $k=1,\dots,n$. Applying this argument recursively shows that setting $\ddot{\lambda_i}(0)=0$ for all $i = 1,\dots,n$ implies $X=0$. In conclusion, if the spectrum of $\Sigma(0)$ is non-degenerate, the geodesic path that joins it with $\Sigma(1)$ cannot preserve all the eigenvalues. 

As a result, the length along the path defined by Eq. \ref{lab:laxpair} will only provide an upper bound on the geodesic length between the stressed and unstressed covariance matrix, or equivalently a lower bound of the plausibility of such a stress test. For practitioners, a lower bound on plausibility may not be very useful as it is difficult to argue in favour of using a stress test which is "not less plausible" than a certain threshold. Of course, one could always find a geodesic joining two covariance matrices which have the same spectrum, but the intermediate covariance matrices along this geodesic will have a different spectrum.

If the spectrum of $\Sigma(t)$ has repeated eigenvalues, the argument shown above still applies to the directions, in the basis of eigenvectors of $\Sigma(0)$, whose eigenvalues have multiplicities of one, so the entries of $X_{ij}$ are zero in these directions. The subset of such matrices $\Sigma$ is however a lower-dimensional subset of the manifold of correlation stress tests, and thus it cannot correspond to a generic correlation stress test that could be applied to an arbitrary covariance matrix.
\section{Example: equities, bonds, and commodities} 
\label{section:fullexample}
Suppose one was presented with an explicit stress test, say a given stressed covariance matrix $\Sigma_2$. By itself,  this would not be a sufficient specification of a covariance stress test, as our approach requires specifying an entire stress path $\Sigma(t)$. However, if a stressed covariance matrix $\Sigma_2$ and and non-stressed covariance matrix $\Sigma_1$ with respect to some preferred basis of risk factors are specified, one can set the stressed path to be the geodesic joining $\Sigma_1$ and $\Sigma_2$. This geodesic is the shortest path joining these two matrices, or equivalently, the path which, at each point, points in the direction of the most plausible stress test. If $\det \Sigma_1 = \det \Sigma_2$, the stress test is a correlation stress, and  one can  calculate
\begin{equation} \label{lab:find_X}
    X = \log{\left(\Sigma_1^{-\frac{1}{2}} \Sigma_2 \Sigma_1^{-\frac{1}{2}} \right)}
\end{equation}
to use in Eq. \ref{lab:stress_test_path} in order to reconstruct the entire path of stress test.

A practitioner may further be faced with a situation where while $\Sigma_1$ is known, the stressed covariance  matrix $\Sigma_2$ is not fully defined, for example only some of its entries are specified. For example, one can be asked to consider returns of bonds, equities, and commodities, and be asked to stress test the correlation of bonds and equities without any specification of what would happen to the other correlations. In such cases, it may be natural to define  the stressed covariance matrix $\Sigma_2$ by requiring that it be the most plausible correlation stress test, namely to find the missing stressed entries of $\Sigma_2$ by the optimisation problem
\begin{equation} \label{lab:argmin}
\Sigma_2 \in \arg \min 	d^2( \Sigma_1, \Sigma_2)
\end{equation}
where $d$ is defined by Eq. \ref{lab:geodesic_distance}.
This may be particularly useful if only a few stressed correlations are specified.

To illustrate the above, Let us return to the stylised example discussed in Section \ref{section:correlation_stress_tests}, 
but with an additional market factor corresponding to commodities with annualised volatility of 25\% and uncorrelated to the other two factors. 
The covariance matrix in the bond--equity basis is thus
\begin{equation} \label{equation:bonds_equities_commodities}
    \Sigma_1 =0.0001 \begin{pmatrix*}[r]
    144 &   0 & 0 \\
      0 &  36 &   0 \\
    0 &   0 & 625 \\
    \end{pmatrix*}.
\end{equation}
Now, stress the correlation of equities and bonds to $0.10$, and that of equities and commodities to $0.20$. What would be the correlation of bonds and commodities? This specification of the stress test could mean:
\begin{equation}
    \Sigma_2 =0.0001 \begin{pmatrix*}[r]
    144 &   7.2  & x \\
    7.2 &   36 &   y \\
    x &     y  &   z \\
    \end{pmatrix*},
\end{equation}
subject to constraints  $0.20 \cdot 12 \sqrt{z}  = x $, $\det \Sigma_1 = \det \Sigma_2$, and $\diag \Sigma_2 > 0 $. 

The values of $x$, $y$, and $z$ that locally minimise the distance defined in Eq. \ref{lab:argmin} are 61.548644, 4.533286, and 657.679793, respectively. Thus, the most plausible result of stressing the equity-bond and equity-commodity correlations as above is
\begin{equation} \label{equation:fitted_matrix}
    \Sigma_2 =0.0001 \begin{pmatrix*}[r]
   144.00 & 7.20 & 61.55 \\ 
  7.20 & 36.00 & 4.53 \\ 
  61.55 & 4.53 & 657.68 \\ 
    \end{pmatrix*},
\end{equation}
to two decimal places. As a result of this most-plausible correlation stress test consistent with equity--bond correlation of 0.10 and equity--commodities correlation of 0.20 the correlation between bonds and commodities becomes 0.0295, perhaps surprisingly low. Note that the volatility of commodities is also affected to keep this a correlation stress, but it increases only marginally to 25.65\%.

The stress generator $X$ in Eq. \ref{lab:stress_test_path} is thus 
\begin{equation} 
    X =\begin{pmatrix*}[r]
 -0.025586 & 0.098662 & 0.201907 \\ 
  0.098662 & -0.005173 & 0.019793 \\ 
  0.201907 & 0.019793 & 0.030759 \\ 
    \end{pmatrix*}.
\end{equation}
While the trace of $X$ may not appear to be zero due to the rounding, it does indeed vanish.  The geodesic path, Eq. \ref{lab:stress_test_path}, can then be extended to an arbitrarily large distance, making the stress progressively less plausible.
\begin{figure}
    \centering
    \includegraphics[width=0.5\textwidth]{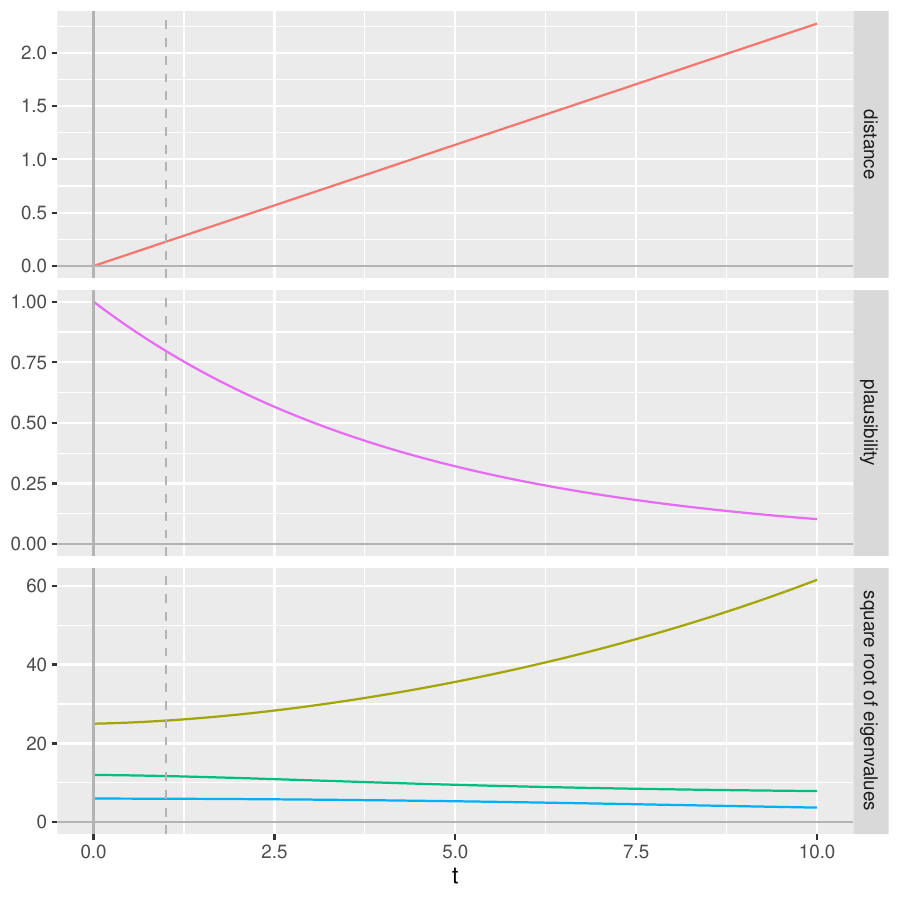}
    \caption{Path of bond--equity-commodity correlation stress tests. The dashed line corresponds to the matrix of Eq. \ref{equation:fitted_matrix}.}
    \label{fig:bondequitypath}
\end{figure}
Figure \ref{fig:bondequitypath} shows the behaviour, as a function of $t$, of the Rao distance, plausibility, and the three eigenvalues of the covariance matrix, extended from the original parameter $t=1$ to $t=10$. Note that the distance is just a linear function of $t$ and the plausibility decreases exponentially with $t$. The smallest eigenvalue (that initially corresponds to the volatility of bonds) tends to zero as $t \rightarrow \infty$. While it may not be apparent from the graph, the differences between the eigenvalues initially increase as a function of $t$, as required by the perturbation formulae. Eqs. \ref{lab:firstOrderCorrection} and \ref{lab:secondOrderCorrection}. The path can be easily extended beyond the range shown in the figure, with numerical accuracy effects becoming non-trivial at $t > 125$, with plausibility becoming less than $4 \times 10^{-10}$ for $t > 120$.

\begin{figure}
    \centering
    \includegraphics[width=1.0\linewidth]{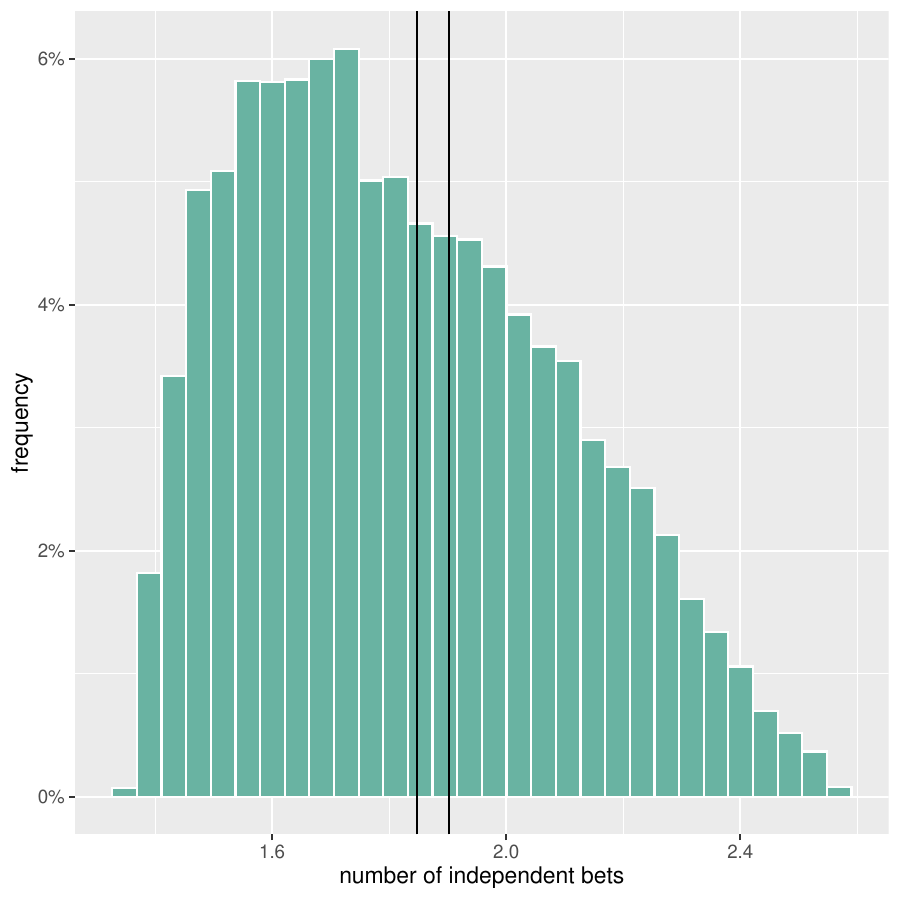}
    \caption{Number of independent bets for a sample of plausibility 0.49 (indicated by the dashed line in Fig. \ref{fig:bondequitypath}). The black vertical lines correspond to 1.85, the number of bets for the fitted matrix given by Eq.\ref{equation:fitted_matrix}, and 1.90, for the unstressed covariance matrix.  }
    \label{fig:ndf}
\end{figure}

\begin{figure}
    \centering
    \includegraphics[width=1.0\linewidth]{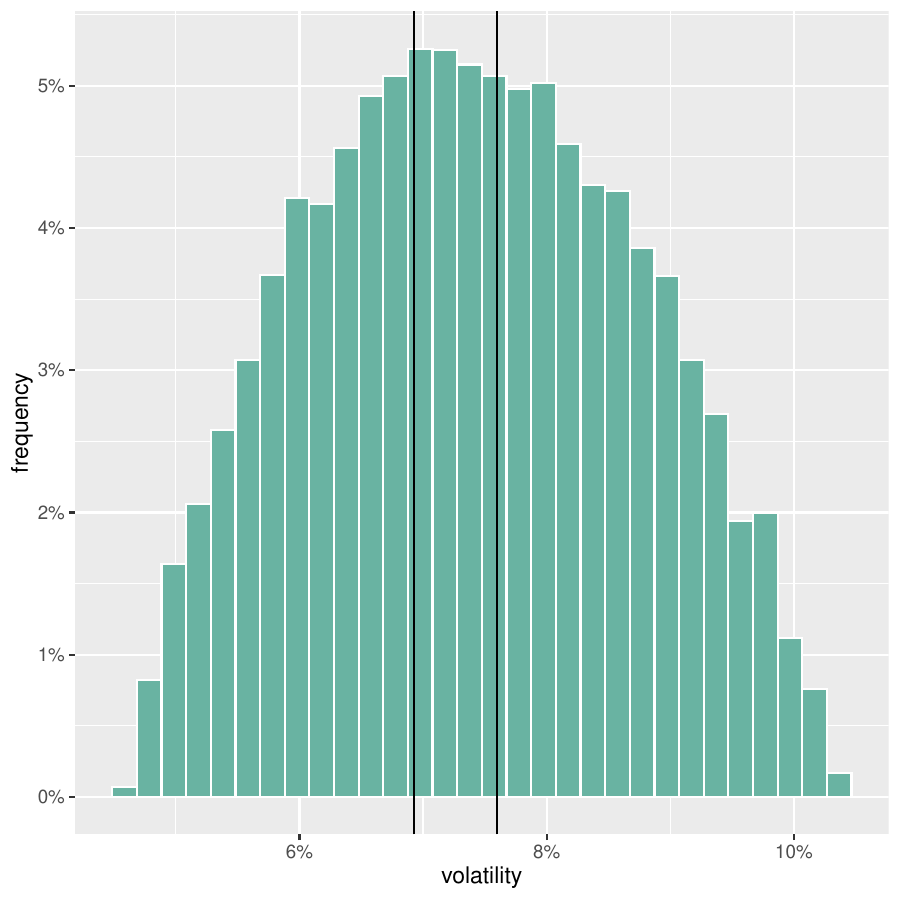}
    \caption{Ex-ante volatility for a sample of plausibility 0.49. The black vertical lines correspond to the volatility of 6.92\% for the unstressed covariance matrix and 7.60\% for the fitted matrix of Eq. \ref{equation:fitted_matrix}}
    \label{fig:volatility}
\end{figure}

 We will now consider a random sample, of size $M=10000$, of uniformly distributed stressed covariance matrices with plausibility $\exp(-1/\sqrt{2}) \approx 0.49$ with respect to the covariance matrix of Eq. \ref{equation:bonds_equities_commodities}. This sample is obtained using the algorithm described in Section \ref{section:sampling}. Figure \ref{fig:ndf} shows the distribution of the effective number of degrees of freedom -- or the exponential of the Shannon entropy of the eigenvalues of $\lambda_i$ of these matrices:
\begin{equation}
    N_{df} = \exp{ \left( - \sum_{i=1}^N p_i \log{p_i} \right) },
\end{equation}
where
\begin{equation}
    p_i = \left( \sum_{i=1}^N \lambda_i \right)^{-1}\lambda_i .
\end{equation}
Figure \ref{fig:volatility} shows the distribution of the ex-ante volatility, calculated using these matrices, for a pro-forma portfolio of 50\% equities, 40\% bonds, and 10\% commodities. One can observe that the ex-ante volatility for a meaningful fraction of such matrices is substantially higher than that of either the unstressed, or the fitted, covariance matrix. The effect would be of course more pronounced for stress tests with lower, but still meaningful, plausibility.

\section{Summary} \label{section:summary}

In this note we have discussed certain geometric properties of the Fisher--Rao manifold that can be used to define covariance and correlation stress tests.

These tests are independent of the non-stressed covariance matrix or the portfolio to which they will ultimately be applied. The correlation stress tests can be extended to have arbitrarily large size. We also discussed a natural measure of the absolute and relative plausibility of the stress tests.

We have proposed to use a simple coordinate-independent parametrisation of correlation stress tests, and discussed
several specific examples: stressing a pair of correlations, stressing all correlations of a single variable, stressing along the diagonal, and stressing all correlations equally.

The proposed correlation stress tests can be used in calculating stressed risk measures such as ex-ante volatility or parametric value-at-risk, factor sensitivities such as betas, incremental risk measures, etc.
For a portfolio of financial assets, such stress tests can be used to assess the values that these risk measures could take reasonably plausible scenarios. For relative-value portfolios, the traditional risk measures sometimes dramatically increase after a market-moving event that results in a meaningful change of the correlations between the positions in the portfolio. The stressed risk measures are potentially less cyclical than the ones traditionally used for limits.  Using  the stressed risk measures for limits may then allow one to maintain and actively manage such relative-value portfolios, without being forced to reduce the positions just because of realised, but unanticipated, changes of correlations. 

The proposed correlation stress tests  can also be applied to analyse stressed valuations of financial derivatives whose price depends on implied correlations, such as spread options, binary options contingent on multiple underlying assets, or  swaps which depend on the volatility of a basket of assets whose individual constituent assets have active option markets. 

\vspace{0.75cm}

I would like to thank Sebastian Armstrong for his comments, in particular for pointing out errors in an earlier version of this note, as well as Juan Antolin-Diaz and the Fulcrum research team for our discussions of this subject. I would also like to thank Witold Biedrzycki for his comments. 
\newpage 
\bibliography{main_article}


 \onecolumngrid

 \newcommand{\PRLsep}{\noindent\makebox[\linewidth]{\resizebox{0.5\linewidth}{1pt}{$\bullet$}}\bigskip}
 \PRLsep
 
Important Risk Information:\\

{\small 
This material is for your information only and is not intended to be used by anyone  other than you. It is directed at professional clients and eligible counterparties only  and is not intended for retail clients. This is not an offer or solicitation with respect to  the purchase or sale of any security. The material is intended only to facilitate your  discussions with Fulcrum Asset Management LLP (“Fulcrum”) as to the opportunities available to our  clients. The given material is subject to change and, although based upon information  which we consider reliable, it is not guaranteed as to accuracy or completeness and it  should not be relied upon as such. The material is not intended to be used as a  general guide to investing, or as a source of any specific investment  recommendations, and makes no implied or express recommendations concerning  the manner in which any client’s account should or would be handled, as appropriate  investment strategies depend upon client’s investment objectives. The price and  value of the investments referred to in this material and the income from them may  go down as well as up and investors may not receive back the amount originally  invested. Past performance is not a guide to future performance. Future returns are  not guaranteed and a loss of principal may occur. Any historical price(s) or value(s) are also  only as of the date indicated. We will endeavour to update on a reasonable basis the  information discussed in this material.

It is the responsibility of any person or persons in possession of this material to  inform themselves of and to observe all applicable laws and regulations of any  relevant jurisdiction. Fulcrum does not provide tax advice to its  clients and all investors are strongly advised to consult with their tax advisors  regarding any potential investment. Opinions expressed are our current opinions as  of the date appearing on this material only.

References to any market or composite indices, benchmarks or other measures of relative  market performance over a specified period of time ("benchmarks") are provided by  Fulcrum for your information purposes only. Investors cannot  invest directly in indices. Indices are typically unmanaged and the figures for the  indices shown herein do not reflect any investment management fees or transaction  expenses. Fulcrum does not give any commitment or undertaking  that the performance or risk profile of your account(s) will equal, exceed or track any  benchmark. The composition of the benchmark may not reflect the manner in which  a portfolio is constructed in relation to expected or achieved returns, portfolio  guidelines, restrictions, sectors, correlations, concentrations, volatility or tracking  error targets, all of which are subject to change over time.

Simulated, modelled, or hypothetical performance results may be used and have certain inherent limitations. Simulated results are hypothetical and do not represent actual trading, and thus may not reflect material economic and market factors, such as liquidity constraints, that may have had an impact on actual decision-making. Simulated results are also achieved through retroactive application of a model designed with the benefit of hindsight. No representation is being made that any client will or is likely to achieve results similar to those shown.

Certain funds managed by Fulcrum Asset Management LLP use quantitative models,  where this is the case, Fulcrum Asset Management LLP can and do make discretionary  decisions on a frequent basis and reserves the right to do so at any point.
Certain transactions, including those involving futures, options and high yield securities  and investments in emerging markets may give rise to substantial risk and may not be  suitable for all investors. Foreign currency denominated investments are subject to  fluctuations in exchange rates that could have an adverse effect on the value or price of,  or income derived from, the investment; such investments are also subject to the  possible imposition of exchange control regulations or other laws or restrictions  applicable to such investments. Investments referred to in this material are not  necessarily available in all jurisdictions, may be illiquid and may not be suitable for all  investors. Investors should consider whether an investment is suitable for their  particular circumstances and seek advice from their investment adviser.

For US Investors: Shares in the Fund will not be registered under the Securities Act of  1933, the securities laws of any state or the securities laws of any other jurisdiction, and  the Fund will not be registered under the Investment Company Act of 1940. There is no  public market for the shares, and no such market is expected to develop in the future.  Shares in the Fund may not be sold or transferred except as permitted under the Fund's  articles of association, and unless they are registered under the Securities Act of 1933 or  an exemption from registration thereunder and under any other applicable securities  law registration requirements is available. Accordingly, investors will be required to bear  the financial risks of an investment in the Fund for an extended period of time.

\textbf{United States:} This document is not intended for distribution to, or use by any person or entity in any  jurisdiction or country where such distribution or use would be contrary to local law or  regulation. In particular this document is not intended for distribution in the United  States or for the account of U.S. persons (as defined in Regulation S under the United  States Securities Act of 1933, as amended (the "Securities Act")) except to persons who  are "qualified purchasers" (as defined in the United States Investment Company Act of  1940, as amended), "accredited investors" (as defined in Rule 501(a) under the  Securities Act) and Qualified Eligible Persons (as defined in Commodity Futures Trading  Commission Regulation 4.7).

\textbf{Switzerland:} The offer and marketing of interests of the fund in Switzerland will be exclusively made to, and directed at, qualified investors (the "Qualified Investors"), as defined in Article 10(3) of the Swiss Collective Investment Schemes Act ("CISA") in conjunction with Art. 4(4) of the Swiss Federal Act on Financial Services ("FinSA"), i.e. institutional clients, at the exclusion of professional clients with opting-out pursuant to Art. 5(3) FinSA (“Excluded Qualified Investors”). Accordingly, the fund has not been and will not be registered with the Swiss Financial Market Supervisory Authority ("FINMA") and no representative or paying agent have been or will be appointed in Switzerland. This document and/or any other offering or marketing materials relating to the interests of the fund may be made available in Switzerland solely to Qualified Investors, at the exclusion of Excluded Qualified Investors

\textbf{Australia:} All information in this communication is intended to be accessed by ‘Wholesale Clients’ as defined by the Corporations Act. No offer or advice: Nothing on the website should be construed as a solicitation, offer or invitation, or recommendation, to acquire or dispose of any investment or to engage in any other transaction. Fulcrum is not providing any personal advice or recommendation regarding any financial products within the meaning of section 766B of the Corporations Act. No consideration has been made of any person’s investment objectives, financial situation and/or needs. Prospective investors should make their own enquiries and should seek all necessary financial, legal, tax and investment advice.

\textbf{Japan:} The Shares have not been and will not be registered pursuant to Article 4, Paragraph 1 of the Financial Instruments and Exchange Law of Japan (Law no. 25 of 1948, as amended) and, accordingly, none of the Shares nor any interest therein may be offered or sold, directly or indirectly, in Japan or to, or for the benefit, of any Japanese person or to others for re-offering or resale, directly or indirectly, in Japan or to any Japanese person except under circumstances which will result in compliance with all applicable laws, regulations and guidelines promulgated by the relevant Japanese governmental and regulatory authorities and in effect at the relevant time. For this purpose, a “Japanese person” means any person resident in Japan, including any corporation or other entity organised under the laws of Japan.

\textbf{For EU Investors:} As at the date of this document, the fund has not been approved, notified or registered in accordance with the Alternative Investment Fund Managers Directive (Directive (2011/61EU) (the “AIFMD”) for marketing to professional investors in any member state of the EEA (each a “Relevant State”). However, such approval may be sought, or such notification or registration may be made in the future. Therefore, this document may only be transmitted to an investor in a a relevant EU Member State at such Investor’s own initiative.

This material has been approved for issue in the United Kingdom solely for the purposes  of Section 21 of the Financial Services and Markets Act 2000 by Fulcrum Asset  Management (“Fulcrum”), Marble Arch House, 66 Seymour Street, London W1H 5BT.
Fulcrum Asset Management LLP is authorised and regulated by the Financial Conduct  Authority (No: 230683).

\noindent \copyright 2026 Fulcrum Asset Management LLP. All rights reserved.

\noindent FC1721 02042026
 }

\end{document}